\documentclass{aa}  
\usepackage{graphicx}
\usepackage{txfonts}
\usepackage{natbib}
\bibpunct{(}{)}{;}{a}{}{,}   
\begin{document}
   \title{$K$-band spectroscopy of IGR~J16358-4726 and IGR~J16393-4643: two new symbiotic X-ray binaries\thanks{Based on observations collected at the European Southern Observatory, Chile (Programme ID  079.D-0668-B)}}


    \author{E. Nespoli
          \inst{1}
          \and
          J. Fabregat\inst{1}
          \and
          R. E. Mennickent\inst{2}
          }

   \offprints{Elisa Nespoli}

   \institute{Observatorio Astron\'omico de la Universidad de Valencia, Calle Catedr\'atico Agust\'in Escardino 7, 46980 Paterna, Valencia, Spain\\
       \email{elisa.nespoli@uv.es}
   \and Departamento de Astronom\'ia, Universidad de Concepci\'on, Casilla 160-C, Concepci\'on, Chile
     }


 
  \abstract
   {Symbiotic X-ray binaries (SyXBs) are a recently discovered subclass of low mass X-ray binaries. Their growing number makes them an important evolutionary channel of X-ray binaries.}
   {Our goal is to perform spectral analysis and classification of the proposed counterparts to IGR~J16358-4726 and IGR~J16393-4643 and establish {their  nature as X-ray systems.}}
   {We used the ESO/UT1 ISAAC spectrograph to observe the proposed counterparts to the two sources, obtaining $K$-band medium resolution spectra ($R = 500$) with a S/N $\gtrsim$ 140. Data reduction was performed with the standard procedure. We classified them by means of  comparison with published atlases. We performed SED fitting to refine the spectral classification.}
   {The two counterparts clearly exhibit the typical features of late-type stars, notably strong CO absorption bands in the red part of the spectrum. With information from previous X-ray studies, we classify the two systems as two new members of the SyXB class. For IGR~J16393-4643, we considered the most probable counterpart to the system, although three other objects cannot be completely discarded. For this system, we compared our findings with available orbital solutions, constraining the orbital parameters and the mass of the companion star.}
   {By including two more systems, we increased to eight the number of known SyXBs, which emerges as a non-negligible category of galactic X-ray binaries.}

   \keywords{X-rays: binaries --- Stars: neutron --- Stars: individual: IGR J16358-4726; IGR J16393-4643
               }

\titlerunning{IGR~J16358-4726 and IGR~J16393-4643: two new SyXBs}
   \maketitle
%

\section{Introduction}

Low-mass X-ray binaries (LMXBs) are interacting systems {consisting of} an accreting compact object and a low-mass (typically $\leq$ 1 $M_{\sun}$) main-sequence or slightly evolved late-type star. A new sub-class of LMXBs has emerged, in which the secondary is an M-type giant. These rare systems are called symbiotic X-ray binaries (SyXBs) after \cite{mas06syxbs} by analogy with symbiotic binaries, in which a white dwarf accretes matter from an M giant companion, either from the wind of the M star or via Roche-lobe overflow. At present, only six X-ray sources are known to belong to this subclass: GX~1+4 \citep{dav77,chak97}, 4U~1700+24 \citep{gar83,mas02}, 4U~1954+319 \citep{mas06syxbs,mat06}, Scutum~X-1 \citep{kap07}, IGR~J16194-2810 \citep{mas07}, and 1RXS~J180431.1-273932 \citep{nuc07}.  \citet[][Table~3]{mas07} presented a summary of the main properties of the first 5 SyXBs, \citet{nuc07} including the last discovered one. 

The increasing number of systems with an evolved giant donor makes this subclass an emerging evolutionary channel of X-ray binaries. They have also been proposed as the probable progenitors of most wide-orbit LMXBs \citep{chak97}.

The six-years of \emph{INTEGRAL}/IBIS data are improving substantially our knowledge about Galactic X-ray binaries \citep[see, for instance,][]{bird07,bir10} by rdetecting a new class of heavily absorbed supergiant massive X-ray binaries \citep[first {proposed} by][]{rev03}, by detecting and allowing the study of supergiant fast X-ray transients \citep[SFXTs; see, for instance,][]{smi06,sguera06,negue07-a,sid07,nes08}, by doubling the number of known high-mass X-ray binaries \citep[HMXBs, see][]{wal06}, and by discovering a vast number of new magnetic cataclysmic variables \citep[CVs; see][]{bar06,mas09}. Among \emph{INTEGRAL} results, one can also identify one new SyXB, IGR~J16194-2810 \citep{mas07}.\\

 \begin{table*}[]
      \caption[]{ISAAC/VLT journal of observations. In the fourth column, we report  the net accumulated exposure time. Column five gives the obtained signal-to-noise ratio. The references list in the last column relates to the identification of the optical/infrared counterpart.}
         \label{table:logobs}
        \centering
            \begin{tabular}{lcccccl}
             \hline
             \hline
             \noalign{\smallskip}
              Source  & K mag &  Start time (UT) &  Exp. time (s) & S/N & IR Counterpart & Reference\\
             \noalign{\smallskip}
              \hline
             \noalign{\smallskip}
            IGR~J16358-4726 & 12.6  & 2007-04-05 08:26   & 480 &  120 &  2MASS~J16355369-4725398& Kouveliotou et al. (2003)\\
            IGR~J16393-4643 & 12.8  & 2007-04-05 08:56   & 1280  & 160 &  2MASS~J16390535-4642137 & Bodaghee et al. (2006)\\
         \noalign{\smallskip}
            \hline
         \end{tabular}
  \end{table*}



IGR~J16358-4726 was discovered by \citet{rev03}, then serendipitously observed by Chandra, which located the source at RA = 16$^h35^m53.8^s$, Dec = -47$^\circ25'41.1''$ with an accuracy of 0.6$''$ \citep{kou03}. This allowed these authors to identify an infrared counterpart, 2MASS~J16355369-4725398. Subsequent astrometry performed by \citet{cha08}, although stating that the 2MASS object is $1.2''$ from the Chandra detection, resulted in establishing it as the true counterpart, mainly because of its brightness ($J$ = 15.41 mag, $H$ = 13.44 mag, $K$ = 12.59 mag).
The X-ray source was identified as a transient, and its spectrum fitted with a power law of spectral index $\Gamma$ $\sim$ 0.5, hydrogen column density $N_\mathrm{H} \sim 3.3\times10^{23}$ cm$^{-2}$, and a FeK$\alpha$ fluorescence line at 6.4 keV \citep{pat07}. From accurate spectral and timing analysis of multi-satellite data, \citet{pat07} also proposed that the source is a neutron star, estimating its magnetic field to be between $10^{13}$ and $10^{15}$ G, if the retrieved spin up was caused by disc accretion, and thus proposing that the object might be a magnetar. A spin period of 5880 $\pm$ 50 s was found by \citet{pat04} and confirmed by several authors \citep{lut05,mer06}. The source is understood to be located in the Norma Galactic arm, at a distance of 5--6 kpc or 12--13 kpc, depending on which crossing of the arm with the line of sight is chosen \citep{lut05}.

From NIR spectroscopy, \citet{cha08} inferred that the system is a HMXB, {and proposed} that the counterpart might be a sgB[e] star because of the detection of H Brackett series, \ion{He}{I}, and \ion{He}{II} absorption lines and a forbidden [FeII] line at 2.22 $\mu$m. This result was then corroborated by \citet{rah08} by means of SED fitting.\\

\citet{sug01} first discovered IGR J16393-4643 with the \emph{ASCA} satellite  and listed it as AX~J16390.4-4642. {During the first \emph{INTEGRAL} scan of the Galactic plane, it was later detected} \citep{mal04}. Subsequent observations with \emph{XMM-Newton}/EPIC derived an improved position (RA = $16^h39^m05.4^s$, DEC = $-46\degr42'12''$), which enabled \citet{bodag06} to identify 2MASS~J16390535-4642137 as the most probable counterpart to the source. 
These authors fitted  the X-ray spectrum with a highly absorbed ($N_\mathrm{H} = 2.5 \times 10^{23}$ cm$^{-2}$) power law ($\Gamma = 0.8 \pm 0.2$), which, with the discovery of a pulse period of 912.0 $\pm$ 0.1 s, led to conclude that the source is an X-ray pulsar \citep{bodag06}.

\citet[][ hereafter T06]{thom06} used \emph{RXTE} data to carry out a pulse timing analysis to determine the orbital parameters of the system. They obtained three mathematical solutions, among which the most plausible one indicated, with an orbital period of 3.7 days and a mass function of 6.5 $\pm 1$ $M_{\sun}$, that the system should be  a HMXB. 
The other two solutions foresaw the following parameters: an orbital period of 50.2 days for a mass function of 0.092  $M_{\sun}$ and an orbital period of 8.1 days for mass function of 221 $M_{\sun}$; both of them were rejected by the authors as being statistically and physically weaker. 
\citet{cha08} obtained accurate astrometry and photometry of the field of IGR J16393-4643, finding three more candidate counterparts, although their lower NIR brightness seemed to favor the 2MASS object. They also performed SED fitting of optical to MIR observations, inferring a BIV-V spectral type for the companion.\\

In this paper, we present $K$-band spectroscopy of the proposed counterparts to the two X-ray systems. In the case of IGR~J16393-4643, our data represents the first available IR spectrum; in the case of IGR~J16358-4726, our data is the first IR spectrum beyond 2.3 $\mu$m. We demonstrate the need for a new classification of the counterparts, which implies a new classification of the systems as SyXBs, if the proposed counterparts are the true ones. In the next section, we describe our observations and data reduction; in Sect. 3, we report the obtained spectra, analyze their features, and propose a classification; in Sect. 4, we discuss our results, and in Sect.~5 we point out the conclusions. Preliminary results of our data analysis were published in \citet{nes08_a}.

\section{Observations and data analysis}    \label{observations}

The two proposed counterparts to the X-ray sources were observed in service mode on 2007 April 5 with the ISAAC spectrograph \citep{moo98} on UT1 at ESO/Paranal observatory. The sky was clear during the observations, the seeing was $\leq1.4''$, and the targets were observed at airmasses of 1.08 and 1.09, respectively. Data were taken in the short wavelength - low resolution mode with a pixel scale of 0.147$''$/pixel and a resolution of 500. Table \ref{table:logobs} reports the observation log, including the achieved signal-to-noise ratio (S/N). Typical on-source integration times for standard stars were between 6 and 10 seconds.\\

Data reduction was performed using the IRAF\footnote{IRAF is distributed by the National Optical Astronomy Observatories which is operated by the Association of Universities for Research in Astronomy, Inc. under contract with the National Science Foundation} package, following standard procedures for IR spectra. We first corrected for the inter-quadrant row cross-talk, a feature that affects the ISAAC detector. {We then performed sky subtraction, applied dome flat-fields}, extracted and rectified the one dimensional spectra. Wavelength calibration was accomplished using xenon and argon lamp spectra. Spurious features, such as cosmic rays or bad pixels, were removed by interpolation, when necessary. The reduced spectra were normalized by dividing them by a fitted polynomial continuum. 

To ensure accurate removal of atmospheric features from the spectra, we followed a strategy similar to that descripted by \citet{clark2000}. Both an \mbox{A0 - A3} \mbox{III-V} and a \mbox{G2-3 V} standard stars were observed immediately before or after each target to obtain very small differences in airmass (differences of 0.008 and 0.007 in airmasses were accomplished for IGR~J16358-4726 and IGR~J16393-4643, respectively). To compute the telluric features in the region of the \ion{H}{I} 21\,661 \AA\ (Brackett-$\gamma$ line, or Br$\gamma$), which is the only non-telluric feature in the A-star spectra, we employed the observed G-star spectra divided by a solar spectrum\footnote{We used the NSO/Kitt Peak FTS solar spectrum, produced by NSF/NOAO} that had been properly degraded in resolution and dispersion corrected with the dispersion solution obtained for the ISAAC spectra. The spectra of the A star, G star, and the solar one were then aligned in wavelength space. A telluric spectrum for each science target was obtained by patching into the A-star spectrum the ratio of the G star to the solar spectrum in the Br$\gamma$ region (we selected the range 21\,590 -- 21\,739 \AA). 
   \begin{figure*}[ht!]
  \centering
   \includegraphics[width=18cm]{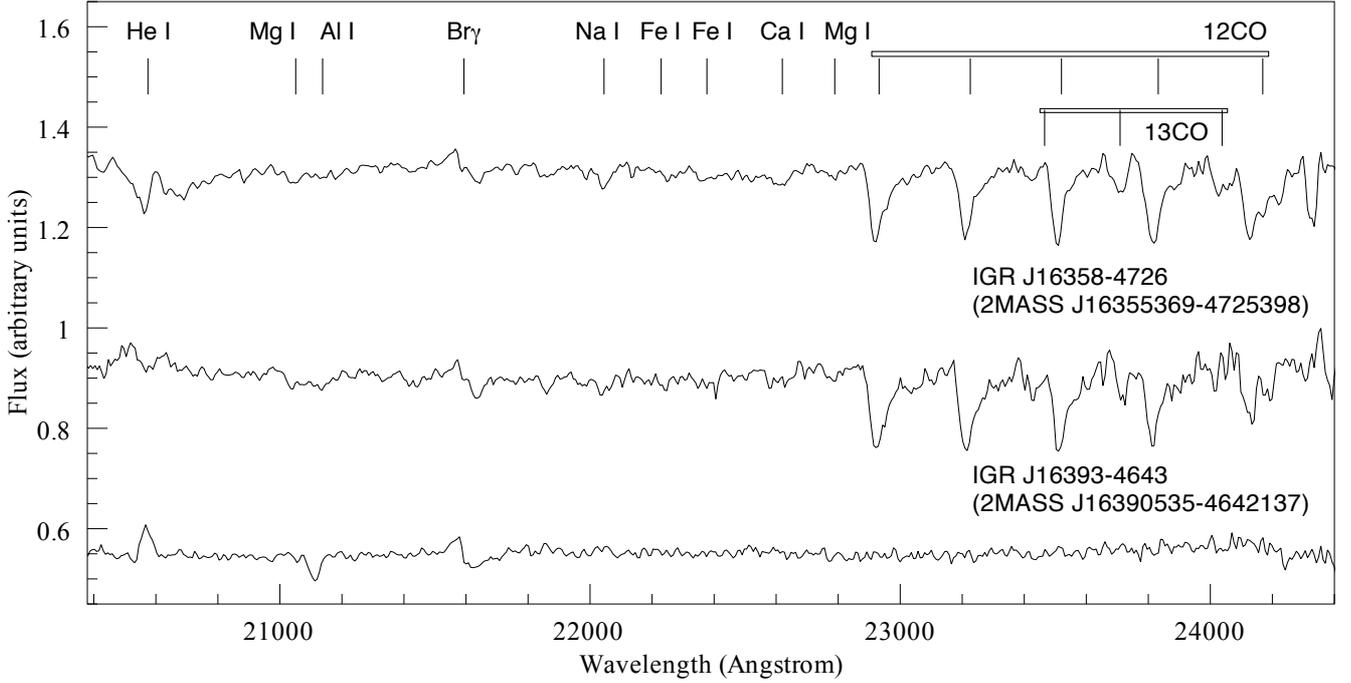}
   \caption{$K$-band spectrum of 2MASS~J16355369-4725398 and 2MASS~J16390535-4642137, the infrared counterparts to IGR~J16358-47262 and IGR~J16393-4643, respectively (first two spectra from above). The positions of identified spectral features are marked by solid lines. With the aim of comparison, we also show the $K$-band spectrum of a blue supergiant (third spectrum from above). The three targets were observed in the same night with the same telescope/instrument configuration. Data reduction was performed in the same way. $^{12}$CO absorption bands ($\lambda \ge 22\,900~ \AA$), typical of late-type stars, are completely absent in the third spectrum. }
             \label{fig:all}
   \end{figure*}
We then corrected for telluric absorption by dividing each science spectrum by its corresponding telluric spectrum. A scale and a shift factor were applied to the calibration telluric spectrum, to correct in the most accurate way for the airmass difference and the possible constant wavelength shift; the optimal values for these parameters were then obtained using an iterative procedure that minimizes the residual noise.

\section{Results}

We present the spectral analysis and classification results for the two targets. The field of NIR spectral classification has {not yet been fully developed} and the level of S/N and resolution required to perform a quantitative profile analysis are very high ($R \sim 12\,000$ and S/N $\gtrsim$ 250). Our analysis is therefore qualitative, based on comparison with available NIR spectral atlases.  \\

The spectra are shown in Fig.~\ref{fig:all} (first two spectra from above), while their most remarkable features are reported in Table~\ref{lines}. Errors in the equivalent widths {are evaluated as the} deviations from their mean values, measured using  different values for the continuum.  
For the sake of comparison, we included in Fig.~\ref{fig:all}  the spectrum of IGR~J16493-4348 (third spectrum from above), a supergiant X-ray binary (SGXRB) that we observed during the same night and conditions as the previous ones. The two first spectra are very similar, their most prominent features being the series of strong CO absorption bands between 2.29 and 2.40 $\mu$m, which are characteristic of late-type stars.
Several metallic lines are clearly recognizable, although at our resolution they are generally blended. Although weaker than the CO bands, they appear significant within the measured errors.\\

 We also tentatively detected \ion{He}{I} 20\,581 \AA\ and Br$\gamma$ in faint emission in both spectra; since they are not a typical spectral feature of late-type stars, these lines might be the signature of an accretion disk around the compact object. However, they are very weak, and we cannot exclude that they might be a residual artifact from the telluric lines removal procedure, which for this region should be performed with great care. This interpretation is consistent with the almost identical profile of the Br$\gamma$ line in the two spectra. 
In any case, emission features typical of X-ray binaries do not seem to be present in almost all of the optical spectra of SyXBs \citep[see][]{mas06syxbs,mas07}. This is because of the high total luminosity of these evolved giant companions, that should overwhelm the light derived from the reprocessing of X-ray irradiation.\\


\begin{table}[!ht]
\caption{{$K$-band line identifications for IGR~J16358-4724, denoted by (I), and IGR~J16393-4643 (II). }}
\begin{center}
\resizebox{8.7cm}{!}{
\begin{tabular}{lccrr}
\hline
\hline
  \noalign{\smallskip}
 Feature           & Transition & Wavel. (\AA)  & EW$_\mathrm{I}$ (\AA)    & EW$_\mathrm{II}$ (\AA)  \\ 
   \noalign{\smallskip}
\hline 
  \noalign{\smallskip}
\ion{He}{I}       & $2s^1S$ -- $2p^1P^o$                                 & 20\,581  &      -1.8$\pm$0.4      &    --~~~~~     \\ 
\ion{Mg}{I}       & $4f^3F_{2,3,4}$ -- $7g^3G^o_{3,4,5}$      & 21\,066    &     0.8$\pm$0.2     &   1.6$\pm$0.3  \\
\ion{Al}{I}       & $4p^2P_{1/2}^o$ -- $5s^2S_{1/2}$      & 21\,099    &         --~~~~~             &       1.8$\pm$0.3 \\
Br$ \gamma$ & 4$^{2}F^{o}$  -- $7^{2}G$                             & 21\,661  &     -2.8$\pm$0.2     &    -0.9$\pm$0.2   \\
\ion{Na}{I}       & $4s^2S_{1/2}$ -- $4p^2P^o_{1/2}$& 22\,100  &                  1.9$\pm$0.4      &     2.0$\pm$0.8\\
\ion{Fe}{I}        & $x^5F_{3}^o$ -- $e^5D_{2}$                         & 22\,263 &     1.6$\pm$0.2      &    0.8$\pm$0.2  \\
\ion{Fe}{I}        & $x^5F_{4}^o$ -- $e^5D_{3}$                         & 22\,387 &     1.2$\pm0.4$      &    1.1$\pm$0.4\\
\ion{Ca}{I}       & $4d^3D_{3,2,1}$ -- $4f^3F^o_{4,3,2}$             & 22\,636& 1.4$\pm0.4$&         1.1$\pm0.4$\\
\ion{Mg}{I}       & $4d^{3}D_{3,2,1}$ -- $6f^{3}F_{2,3,4}^o$   & 22\,800 &     2.7$\pm$0.5     &     1.6$\pm$0.4  \\
$^{12}$CO      & (2,~0) bandhead                                              & 22\,900 &    9.9$\pm$0.3       &     11.9$\pm$0.4\\
$^{12}$CO      & (3,~1) bandhead                                              & 23\,226 &    13.0$\pm$0.6    &     12.9$\pm$0.6\\
$^{13}$CO      & (2,~0) bandhead                                              & 23\,448 &    2.5$\pm$0.4       &     2.5$\pm$0.8\\
$^{12}$CO      & (4,~2) bandhead                                              & 23\,524 &    7.8$\pm$0.4      &     9.8$\pm$0.3\\
$^{13}$CO      & (3,~1) bandhead                                              & 23\,739 &    4.2$\pm$0.2       &      4.5$\pm$0.8\\
$^{12}$CO      & (5,~3) bandhead                                              & 23\,829 &    12.8$\pm$0.6    &      8.05$\pm$0.7\\
$^{13}$CO      & (4,~2) bandhead                                              & 24\,037 &    3.6$\pm$0.5       &      1.98$\pm$0.3\\
$^{12}$CO      & (6,~4) bandhead                                              & 24\,156 &    10.1$\pm$0.8    &     7.0$\pm$0.4\\
  \noalign{\smallskip}
  \hline
\end{tabular}}
\end{center}
\label{lines}
\end{table}


Our ISAAC spectra clearly infer that both objects are of late-type and exclude that the systems are HMXBs with OB companions. From the identified spectral lines, our conclusion is that both stars are of K or M spectral type \citep[see, for instance,][for a comparison with atlases of cold stars]{klei86,for00,iva04}. The forbidden [Fe II] transition at 22\,200 Å detected in IGR~J16358-4726 by \citet{cha08} is not present in our spectrum. The general morphology of the spectra and the comparison between the relative strength of CO 22\,900 \AA\ and both \ion{Na}{I} and \ion{Ca}{I} lines, which are luminosity indicators \citep{ram97}, indicate that both stars are giants. In particular, from the empirical relation EW(CO) versus.~$T_\mathrm{eff}$ that these authors found, our EW($^{12}$CO(2,~0))  infers a K giant classification for both objects, and nominal values of $T_\mathrm{eff} \sim 4400$ K for IGR~J16358-4726 and $T_\mathrm{eff} \sim 4250$ K for IGR~J16393-4643.

We believe that the level of precision achievable in our spectral classification cannot be higher than this, because of the lack of reliable $T_\mathrm{eff}$  indicators in the $K$-band in this spectral type range. Adequate indicators in the NIR domain are present in the $J$ and $H$ bands \citep{wri08}. Unfortunately, we have no spectra for these ranges.\\

To corroborate the accuracy of our analysis and to avoid any possible mistaking of CO  absorption bands with telluric components, we included in Fig.~\ref{fig:all}, besides the spectra from the two sources described in this work, the spectrum of IGR~J16493-4348, a SGXRB that we observed in the same night and conditions as the previous ones.
We classified  the counterpart to IGR~J16493-4348 as a B0.5Ib star \citep{nes08_b} and, as one can see, the telluric corrected spectrum of the source shows no evidence of the deep absorption features that characterize the spectra that are the subject of this work.\\

\begin{figure}[t]
   \centering
\includegraphics[bb= 0 10 382 248, clip, width=8.5cm]{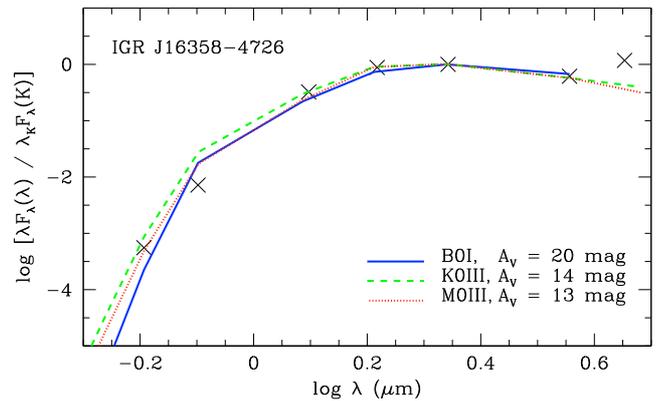}
   \caption{Spectral energy distribution of IGR~J16358-4724 (crosses) and SED obtained from intrinsic colors for three different spectral types (lines).The errors for the photometric data are within the size of the marks.}
              \label{fig:sed16358}
    \end{figure}

\begin{figure}[!h]
   \centering
   \includegraphics[bb= 0 10 382 248, clip, width=8.5cm]{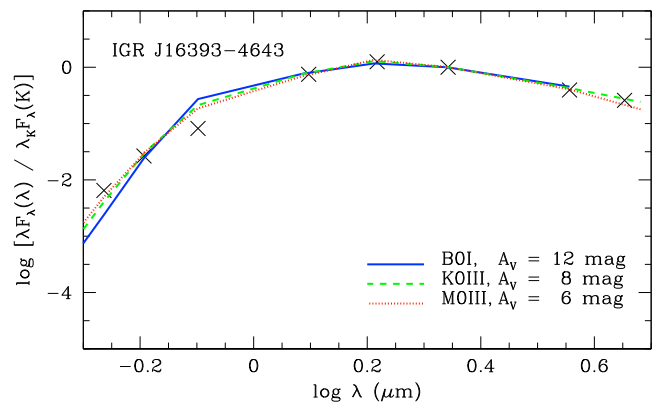}
   \caption{The same as Fig.~\ref{fig:sed16358}, for IGR~J16393-4643. The errors in the photometric data are within the size of the plot symbols.}
              \label{fig:sed16393}
    \end{figure}

 In Figs.~\ref{fig:sed16358} and \ref{fig:sed16393}, we present the spectral energy distribution  of IGR~J16358-4724 and IGR~J16393-4643,   respectively, in the log$(\lambda F_{\lambda}(\lambda) / \lambda_{K}F_{\lambda}(K))$ -- log $\lambda$ plane, using photometry given by \citet{cha08} and \citet{rah08}. This includes photometry in the bands $RIJHK$ and the GLIMPSE 3.6 $\mu$m and 4.5 $\mu$m bands for both sources; for IGR~J16393-4643, $V$ photometry was also available. We built SEDs corresponding to different spectral types, according to the intrinsic colors given by \citet{pic98} for optical/NIR bands and by \citet{koo83} for MIR. In the figures, data are marked by crosses, and fitted SEDs by lines. Photometric data and spectral type distributions are normalized in the $K$ band. 
We have did not attempt formal fitting procedures. The aim of the figures is to illustrate that very different choices of spectral type and reddening produce reasonable fits to the data. In our calculation, we employed the mean extinction law, $R_{V}= 3.1$ \citep{rieke85}.\\

We plot a blue supergiant and two red giants as representative examples {of possible spectral types}. We found that, even with slight differences, every spectral type can be fitted to the data by choosing the appropriate value of extinction. In the case of IGR~J16358-4726, we obtained best fits by choosing $A_{V} = 20$ mag for a B0I type, $A_{V} = 14$ mag for a K0III type, and $A_{V}=13$ mag for a M0III type. Analogous results were obtained for IGR~J16393-4643 with $A_{V} = 12 $ for a B0I, $A_{V} = 8 $ for a K0III and $A_{V} = 6$ for a M0III.

The reason why very different spectral types are able to reasonably fit the data {is} the degeneracy between the direction of the reddening vector and the effective temperature variation in the spectral range that we employed, which makes almost every spectral type possible if the opportune extinction value is chosen. The results provided by SED fitting do not therefore allow the spectral classification to be improved, but provide a valuable tool with which to calculate the interstellar extinction of any assumed spectral type.\\

In the case of IGR~J16393-4643, the low values obtained for the extinction are inconsistent with the high hydrogen column density measured in the X-ray range, namely \mbox{$N_\mathrm{H} = 2.5 \times 10^{23}$ cm$^{-2}$}. This may indicate that the star is not the true counterpart to the system; another explanation could be an additional source of extinction intrinsic to the system, which only affects the compact object where the X-ray emission originates. Enormous amounts of absorption, with $N_\mathrm{H} \sim 10^{23}$ cm$^{-2}$, has been observed for other SyXBs, \citep[see, for instance,][]{mat06}. \\

For the aforementioned spectral types and corresponding extinction values, we calculated the distance using the relation $M_{K} = K + 5 - 5$ log $d - A_{K}$. In each case, absolute magnitudes and intrinsic color indices were obtained from \citet{pic98}.  The estimated distance values were, for IGR~J16358-4726:  
36.6 kpc (if B0I), 1.4 kpc (if K0III), and 20.1 kpc (if M0III); for IGR~J16393-4643: 24.1 kpc (if B0I), 1.0 kpc (if K0III), and 13.9 kpc (if M0III).  For both sources, a type B0I counterpart would thus provide an unrealistically large estimate of the distance; the same occurs, for IGR~J16358-4726 only, in the case of a M0III companion.\\

The remaining estimates are compatible with both the classical \citep[four arms,][]{val08} and the more recently proposed \citep[two main arms,][]{chu09} spiral structure models of our Galaxy.
A K0I companion would locate both sources in the Sagittarius/Carina inner arm according to the classical model, or the Sagittarius minor arm, according to the most recently proposed model. Finally, a M0III companion for IGR~J16393-4643 would imply that the source is placed in the tangential region of the Norma arm with both models. \\

The error estimation for the distance depends mainly on the error in the mean absolute magnitude assigned to the spectral type, which can be very large and reach more than 50$\%$ when propagated to the distance calculation.   
Since the derived extinction value for each type would also apply to a wide range of subtypes, the consequent distance value range would largely extend. The distance inferred should thus be interpreted with care.\\

\section{Discussion}
On the basis of the spectral classification of the proposed counterparts as cold giants/supergiants and the X-ray features of the systems illustrated in Sect.~1, we propose that both of them belong to the class of SyXBs,  in which the companion to the neutron star is an evolved late-type star. In the discussion that follows, we use the terms giant and supergiant in its physical sense, i.e., related to the mass, internal structure, and evolutionary status of the star and not to its luminosity class. As discussed above, a precise determination of the spectral type and luminosity class is difficult using  our K-band spectra.\\

A red giant represents the late evolutionary stage of either a low-mass star (0.8 to 2.3 $M_{\sun}$) or an intermediate-mass star (2.3 to 10 $M_{\sun}$). The red giant phase includes the shell hydrogen-burning phase (first giant branch), the core helium-burning phase (horizontal branch or ``red clump'' for low-mass stars, ``blue loop'' for intermediate-mass stars), and the shell helium-burning phase (asymptotic giant branch). We refer, for instance, to \citet{ibe74} and \citet{ibe83}. Red supergiants are short-lived stages in the late evolution of moderately massive stars \citep[10 to 25 $M_{\sun}$, see][]{mass09,fig06}\\

In the case of IGR~J16358-4724, we classified the proposed counterpart as a K-M giant/supergiant. Strong support for this system hosting a neutron star is provided by the detection of X-ray pulsations \citep{pat04} in addition to the description of the X-ray energy spectrum as an absorbed power law model, a high energy cut-off, and an Fe line \citep{pat07}, which are typical features of neutron stars in Galactic X-ray binaries. The designation of 2MASS~J16355369-4725398 as a counterpart was confirmed in the literature \citep[see][]{kou03, cha08}. Previous NIR spectroscopy from \citet{cha08} was used to classify it as a sgB[e], but, as the authors warned, their NIR spectrum was very faint and extremely absorbed. By ending at 2.3 $\mu$m, it also missed the last and most significant portion of the $K$ band, where CO bands clearly indicate the K-M spectral type of the star.
We propose that these factors could lead to a misclassification of the system.\\

In the case of IGR~J16393-4643, our spectrum is also indicative of a K-M giant/supergiant. The presence of a pulsar in the binary system is validated by both the spectral shape of the X-ray emission \citep{bodag06} and the detection of pulsations (T06). Because of its magnitude, the optical counterpart that we considered is the most probable one among the four objects within the error circle of the X-ray detection.\\

According to T06, IGR~J16393-4643 has an orbital period of 3.7 days and a mass function of 6.5 $M_{\sun}$ (orbital solution 1 in T06). It seems unlikely that an intermediate-mass  red giant can fit in such a narrow orbit. To check this, we use the physical parameters of a 7 $M_{\sun}$ star along its evolutionary track given by \citet{scha92}.\\

The size of the Roche lobe of a star in a binary system is usually characterized by the \emph{Roche radius} $R_\mathrm{L}$, defined as the radius of a sphere that has the same volume as the Roche lobe. To calculate the Roche radius, we used the \citet{eggl83} approximation


  \begin{equation}
    \frac{R_L}{a} = \frac{0.49~ q^{2/3} }{0.6~ q^{2/3}+~ \mathrm{ln}~(1 + q^{1/3})}~, 
\label{roche}
  \end{equation}

\noindent where $a$ is the semimajor axis and $q$ is the mass ratio of the binary components ($q = M_\mathrm{c} / M_X$).
We used the common approximation for the mass of a neutron star, $M_X = 1.4$ $M_{\sun}$.

To obtain the Roche radius for solution 1 of T06, we first calculated the major axis of the orbit from the third Kepler law

\begin{equation}
M_X + M_\mathrm{c} = \frac{4\pi^2}{G}\frac{a^3}{p^2}~,  
\label{kepl}
\end{equation}

\noindent 
where $G$ is the gravitational constant that we wrote as \mbox{$G = 2945.02~  R_{\sun}^3 ~M_{\sun}^{-1}$ d$^{-2}$}, $a$ is the semimajor axis, and \mbox{$p = 3.7$ d} is the orbital period. 
For the above parameters, we obtained a Roche radius of $R_L = 10.68$ $R_{\sun}$.

According to \citet{scha92}, the radius of a 7 $M_{\sun}$ star at the beginning of the giant branch is  86.5 $R_{\sun}$. {During its later evolution, at the core He-burning phase}, the minimum radius reached is 50 $R_{\sun}$. The only evolutionary phase in which a 7 $M_{\sun}$ star can fit wihin a 10.68 $R_{\sun}$ Roche lobe is during the main sequence, for temperatures higher than 15\,000 K. This  corresponds to a B-type spectrum, which is completely rejected by our spectral classification. This result shows that the system IGR~J16393-4643, in the binary orbit proposed by T06, cannot harbor a late-type giant star of $\sim$7 $M_{\sun}$.\\
 


Among the orbital solutions calculated by T06, the above discussion thus excludes solution 1. Our conclusion may be supported by the non-detection of any orbital period of the order of a few days (as solution 1 estimates) by \citet{bodag06} from ISGRI light curves. We also exclude solution 3 because of, as the authors point out, its statistical weakness and physically unrealistic constraints. Although statistically less favored than the first, solution 2 appears to be the only acceptable solution. With a mass function of 0.092 $M_{\sun}$ and an orbital period of \mbox{50.2 days}, it is indeed compatible with our spectral classification, as we now show.\\

To constrain the mass of the companion, we calculated, for increasing values of mass, the corresponding radius at the beginning of the giant branch, and compared it with the maximum permitted one, \emph{i.e.}, the Roche radius. \\

We again we first calculated the major axis of the orbit by applying the third Kepler law in Eq.~(\ref{kepl}), using \mbox{$p = 50.2$ d} from solution 2 as the orbital period.


\begin{table}[!t]
\caption{Results from the analysis of the orbital parameters of IGR~J16393-4643 for different values of mass of the companion star $M_\mathrm{c}$. Tabulated values for $T_\mathrm{eff}$ and $L$ are taken from Schaller et al. 1992, for a star at the beginning of the giant branch. For each case, the last column shows the corresponding spectral type.}
\begin{center}
\resizebox{9cm}{!}{
\begin{tabular}{cccccccc}
\hline
\hline
  \noalign{\smallskip}
 $M_\mathrm{c}$           & $a$  & $R_L$   & log $T_\mathrm{eff}$   &   log($L/L_{\sun}$) & $R_\mathrm{c}$  &  $i$  & Sp.~t. \\ 
($M_{\sun}$) &  ($R_{\sun}$) & ($R_{\sun}$) & & &($R_{\sun}$)& ($^{\circ}$)  & \\
   \noalign{\smallskip}
\hline 
  \noalign{\smallskip}
1 & 76.5  &  26.8  & 3.72 &  0.33 &  1.75 & 54.0 & G5 III  \\
2 & 85.9 & 35.2   & 3.71 & 1.23 & 5.19   & 30.7 &  G5 III \\
3 & 93.6 & 41.7   & 3.70 & 1.95 & 12.46 & 23.8   & G8 III\\
4 & 100.3 & 47.3 & 3.68 & 2.46 & 24.35  &  20.3   & K0 III\\
5 & 106.1 & 52.1 & 3.67 & 2.85 & 40.93  & 18.1  &  K1 III\\
\noalign{\smallskip}
\hline
\end{tabular}}
\end{center}
\label{tab:rad}
\end{table}



Taking the common approximation $M_X = 1.4$ $M_{\sun}$, we solved E.~\ref{kepl} with respect to $a$ for different values of $M_{c}$, and employed the retrieved value in Eq.~\ref{roche} to obtain the corresponding Roche radius. We then considered a red giant at the beginning of the giant branch, took tabulated values of $T_\mathrm{eff}$ and $L$ from \citet{scha92}, and solved the black-body equation, $L=4\pi R^{2}\sigma T_\mathrm{eff}^{4}$, to obtain the stellar radius. For each value of mass, starting with $M_\mathrm{c} = 1$ $M_{\sun}$, we compared the  retrieved radius with the Roche radius of the system. Results from this analysis are reported in Table~\ref{tab:rad}, where we also show, for each case, the corresponding spectral type, obtained from temperature and luminosity, and the inclination $i$ of the system as calculated from the mass function of solution 2.\\

We emphasize that the parameters $R_\mathrm{c}$ and $i$ in Table~\ref{tab:rad} correspond to the beginning of the first giant branch ascent. In consequence, they only represent a lower limit to the radius and upper limit to the inclination angle for stars of the given mass.\\

Up to $M_\mathrm{c} = 5$ $M_{\sun}$, a red giant is able to fit within the orbit foreseen by solution 2 of T06, while for higher values the stellar radius would exceed the Roche radius of the system. We thus set an upper limit of 5 $M_{\sun}$ to the mass of the companion star. This mass limit excludes the possibility that the companion is a  supergiant since the lower limit to the mass of a red supergiant is 10 $M_{\sun}$ as discussed above.\\

An alternative explanation is that 2MASS~J16390535-4642137 is not the true optical/infrared counterpart to IGR~J16393-4643. A spectroscopic study of the other three stars \citep[see][Fig.~2c]{cha08} found in the error box  of the X-ray source is needed to confirm or exclude this case. If one of those targets were an OB supergiant, this would then support the classification of the system as a HMXB, and would make the orbital solution 1 from T06 acceptable.\\

For the sake of comparison, we collect in Table~\ref{tab:properties} the main properties of the two systems studied, thus extending Table~3 by \citet{mas07}. The mass accretion rate was calculated from the X-ray luminosity, as 

\begin{equation}
\dot{M} = \frac{L_\mathrm{X} R}{G M}~,
\label{accr_rate}
\end{equation}

\noindent
where $R\sim 10$ km is the radius of the neutron star, and $M \sim 1.4 M_{\sun}$ its mass. For IGR~J16393-4643, $L_\mathrm{X}$ was obtained from the X-ray 2--10 keV flux given by \citet{bodag06}, and assuming a distance of  1.4 kpc, according to our estimate in the most probable case of a K0III counterpart.

Although  the class of SyXBs is not yet well characterized we note that the known properties of these two systems and those of the other six members are homogeneous. We confirm the peculiarity of the class with respect to the spin, which exhibits periodicities spanning a wide range, from hundreds of seconds to hours \citep[see][ and references therein]{mas07}.


\begin{table}[!t]
\caption{Synoptic table reporting the main known parameters for IGR~J16358-4724 and IGR~J16393-4643. The X-ray luminosity is given in the 2--10 keV band. The mass accretion rate $\dot{M}$ was estimated assuming typical parameters for a neutron star, \emph{i.e.} $R_\mathrm{X} = 10$ km and $M_\mathrm{X} = 1.4$ $M_{\sun}$.  }
\begin{center}
\resizebox{9cm}{!}{
\begin{tabular}{l | cc}
\hline
\hline
  \noalign{\smallskip}
Parameter & IGR~J16358-4724  & IGR~J16393-4643\\
   \noalign{\smallskip}
\hline 
  \noalign{\smallskip}
Spectral type &  &   \\
of the secondary & K-M III [1]& K-M III [1] \\[1ex]
$V$-band magnitude & $>23.67$ [2] & 21.53 [2]\\[1ex]
$A_{V}$ (mag)  &  12--13 [1] & 4--6 [1]  \\[1ex]
Distance (kpc)  & 5--6 ; 12--13 [3] & $\sim$ 10 [4] \\[1ex]
X-ray spectrum &  [5]  &  [4] \\[1ex]
$L_\mathrm{X}$  (erg s$^{-1}$) & $3 \times 10^{32}$ -- $2 \times 20^{36}$ [5] & ${2\times 10^{34}}$ [4,1]\\[1ex]
$L_\mathrm{X} / L_{secondary}$ & --   & -- \\[1ex]
$\dot{M}$ (g s$^{-1}$) & $1.6 \times 10^{12}$ -- $1.1 \times 10^{16}$ [1]  & ${1.1\times10^{14}}$ [1]\\[1ex]
$P_{spin}$ (s)  &   5850 [6] & $\sim 912$ [4,7]\\[1ex]
$\dot{P}_{spin} / P_{spin} $ (s$^{-1}$)   & 3.1 $\times 10^{-8}$ [5] & 1.0 $\times 10^{-11}$ ? [7,1]\\[1ex]
$P_{\mathrm{orb}}$ (d) & -- &  50.2 ? [7,1]\\
\noalign{\smallskip}
\hline
  \noalign{\smallskip}
\multicolumn{3}{ l }{ References: }\\
\multicolumn{3}{l}{[1] this work; [2] Chaty et al. (2008); [3] Lutovinov et al. (2005);}\\ 
\multicolumn{3}{ l }{ [4] Bodaghee at al. (2006); [5] Patel at al. (2007); [6] Patel et al. (2004); }\\
\multicolumn{3}{ l }{[7] Thompson et al. (2006).}\\
  \noalign{\smallskip}
\hline
\end{tabular}}
\end{center}
\label{tab:properties}
\end{table}


\section{Conclusions}

We have used infrared medium-resolution spectroscopy to classify the proposed counterparts to IGR~J16358-4724 and IGR~J16393-4643. In this work, we have found that:

\begin{itemize}
\item [-] both objects exhibit the typical features of cold giant or supergiant stars. Our classification constrains the spectral type to be K-M.
\item [-] previous classifications of the systems as HMXBs have been dismissed.
\item [-] spectral classifications of the counterparts as late type-giant stars, and an X-ray behavior characteristic of neutron stars known from previous works, has allowed us to include both systems in the SyXB class, increasing the number of this group from six to eight members. 
\item [-] among the orbital solutions for IGR~J16393-4643 proposed by \citet{thom06} from a pulse-timing analysis, only one, {with  a mass function of 0.092 $M_{\sun}$ and orbital period of 50.2 days}, is compatible with our spectral classification.
\item[-] an upper limit to the mass of the companion star in IGR~J16393-4643 was set to be 5 $M_{\sun}$.
\end{itemize}

The results that we have obtained for IGR~J16393-4643 are valid if the candidate counterpart is confirmed by excluding as a possible companion one of the other three sources found in the error circle of the \emph{{XMM-Newton}} position. Future spectroscopic studies of these faint sources will be able  to confirm without any doubt the classification of the system.  
Our data clearly highlight the importance of obtaining $K$-band spectroscopy beyond 2.3 $\mu$m, where CO absorption features are located. With spectra being limited shortward of this range, which can be featureless in particular at low resolution and S/N, classification can be difficult to perform and misleading. 

  \begin{acknowledgements}
The work of EN and JF is supported by the Spanish Ministerio de Educaci\'on y Ciencia, and FEDER, under contract AYA 2007-62487. This work has been partly supported by the Generalitat Valenciana project of excellence PROMETEO/2009/064. EN acknowledges a ``V Segles'' research grant from the University of Valencia. REM acknowledges support by Fondecyt 1070705, the Chilean Center for Astrophysics FONDAP 15010003 and the BASAL Centro de Astrof\'isica y Tecnolog\'ias Afines (CATA) PFB--06/2007. 
  
      \end{acknowledgements}

\bibliographystyle{aa}

\end{document}